\documentclass[12pt]{article}
\usepackage{latexsym}
\usepackage{geometry} % see geometry.pdf on how to lay out the page. There's lots.
\usepackage{pict2e}

\def\bs{\begin{subequations}}
\def\es{\end{subequations}}

\catcode`\@=11

\newtoks\@stequation
\def\subequations{\refstepcounter{equation}
  \edef\@savedequation{\the\c@equation}%
  \@stequation=\expandafter{\theequation}%   %only want \theequation
  \edef\@savedtheequation{\the\@stequation}% % expanded once
  \edef\oldtheequation{\theequation}%
  \setcounter{equation}{0}%
  \def\theequation{\oldtheequation\alph{equation}}}

\def\endsubequations{\setcounter{equation}{\@savedequation}%
  \@stequation=\expandafter{\@savedtheequation}%
  \edef\theequation{\the\@stequation}\global\@ignoretrue}

\makeatletter%  allow access to internal LaTeX commands
        \renewcommand{\theequation}{\thesection.\arabic{equation}}%
        \@addtoreset{equation}{section}%
\makeatother%  turn off access to internal LaTeX commands

\usepackage{hyperref}

\begin{document}
\begin{titlepage}
\begin{center}
 {\bf {Tachyons with any Spin} }
 
 Charles Schwartz, Department of Physics \\University of California, Berkeley, CA. \\
 schwartz@physics.berkeley.edu   \\
 final version \hspace{.5in}        Revised December 3, 2021  \\
 
  \end{center}
\vskip 1cm
 accepted by IJMPA 
\vfill

\begin{abstract}
In earlier work we showed how to handle the Group Theoretical issue of the Little Group for spin 1/2 tachyons by introducing a special metric in the vector space of one-particle states. Here that technique is extended to tachyons of any spin. Examining the bi-linear algebra of the generating matrices for  spin 5/2, we find a  complete basis for the Gell-Mann matrices that form the Lie algebra for SU(3). A Dirac-like equation is developed for tachyons of any integer-plus-one-half spin; and it shows multiple distinct mass eigenvalues. The primary model shows a mass spectrum (in the case of j = 5/2) that roughly mimics the known data on masses of the three neutrinos; the model can be tweaked to fit that experimental data precisely.
\end{abstract}

\vfill

\end{titlepage}

\section{Introduction}
Traditionally, the Group Theoretical approach to a quantum theory of elementary particles leads to the conclusion that tachyons (faster than light particles) can exist only with spin zero. That mathematical approach involves finding the appropriate Little Group, involving three members of the Lie algebra for the Lorentz group. For ordinary (slower than light) particles this Little group is identified with O(3); but for tachyons it is O(2,1) and this does not allow \emph{finite dimensional} unitary representations, except for the one-dimensional case.

In earlier work \cite{CS1} I showed how to overcome that problem by the introduction of a particular indefinite metric in the space of one-particle states - for the case of spin 1/2. That is reviewed in Section 2.  Some people worry that introducing an indefinite metric can lead to states with zero norm. To avoid such problems we can assert a selection rule that says the physical states are eigenstates of the helicity. Constructing a full theory of tachyons with interactions is a future task and here we are just studying free particles. Nevertheless, this comment serves to elevate  the idea that neutrinos are the most likely candidates to be tachyons.

In Section 3 I will show how to generalize that earlier result for any spin; and then, in Section 5, examining the special case of $j=5/2$, we are led to some delightful algebraic discoveries that connect to the famous SU(3) group. Section 6 turns to the development of a Dirac-like equation for tachyons of any spin; and we are finally led to a successful formulation that has multiple mass states. In an Appendix it is shown that the primary model gives a mass spectrum (in the case of j = 5/2) that roughly mimics the known data on masses of the three neutrinos; the model can be tweaked to fit that experimental data precisely.

My conventions are that the Minkowski metric has $\eta_{00} = +1$ and the velocity of light is c=1.
  
 \section{The Little Group for Tachyons}
 Start with the assumption of a one-particle state $|p^\mu, \alpha>$ identified by a momentum 4-vector with the constraint $p^\mu p_\mu = - M^2$, and some other (internal) quantum numbers $\alpha$. We now make a Lorentz transformation to a specially selected frame of reference (for ordinary particles this would be the "rest frame") in which the spacelike four-momentum has the form $p_*^\mu = (0,0,0,M)$ (this is the frame in which the tachyon has infinite velocity, in the z-direction).
 
 The full Lorentz group is generated by the six operators $J_i, K_i, i = 1,2,3$. One can readily verify that the three operators $J_3, K_1, K_2$ leave that vector $p_*^\mu$ unchanged. These define the Little Group; and their Lie algebra is,
 \begin{equation}
 [J_3, K_1] = iK_2, \;\;\;[J_3,K_2] = -i K_1, \;\;\; [K_1, K_2] = -i J_3.\label{b1}
 \end{equation}
 That minus sign in the last equation tells us that this is not the rotation group O(3) but the (so-called non-compact) group O(2,1). Let us see why this is a problem for physics.
 
 For the O(3) group, generated by the three J's, one can write the general group element as
 \begin{equation}
 S(3) = exp(i\alpha J_1 + i\beta J_2 + i\gamma J_3) ,\label{b2}
 \end{equation}
 where $\alpha, \beta, \gamma$ are three arbitrary real constants; and one sees that this is a unitary operator $S^\dagger = S^{-1}$ because all the generators are Hermitian operators.
 
 Looking at the algebra for the O(2,1) Little group, we see that if we replace $K_1 \rightarrow i J_1$ and $K_2 \rightarrow iJ_2$, then this looks exactly like the Lie algebra for O(3). So, we can write the general group element for O(2,1) as,
 \begin{equation}
 S(2,1) =exp(-\alpha J_1 - \beta J_2 + i\gamma J_3),\label{b3}
 \end{equation}
 where these J's are any (Hermitian) representation of O(3). This S(2,1) is not a unitary operator. This is a problem in that we traditionally want to define a vector space for one-particle states where Lorentz transformations are represented as unitary operators in any inner product.
 \begin{equation}
 <\psi |\phi> \longrightarrow <S\psi| S\phi> = <\psi|S^\dagger S|\phi> = <\psi|\phi>.\label{b4}
 \end{equation}
 
 The resolution of this problem, as given in my earlier work \cite{CS1}, is to introduce a particular metric, which I call H, into this space of one-particle states.
  \begin{equation}
 <\psi |H|\phi> \longrightarrow <S\psi|H| S\phi> = <\psi|S^\dagger HS|\phi> = <\psi|H|\phi>.\label{b5}
 \end{equation}
 The requirement is that  $S^\dagger H = H S^{-1}$. In terms of the algebra, this requires that
 \begin{equation}
 HJ_3 = J_3 H, \;\;\; HJ_1 = - J_1 H, \;\;\; HJ_2 = -J_2 H. \label{b6}
 \end{equation}
 For the case of spin 1/2, represented by the 2-dimensional Pauli matrices, we readily found the solution: $H = \sigma_3$.
 
 \section{Generalize to any spin} 
 We are all very familiar with the standard matrix representations for the generators of rotation group O(3).
 The matrices $J_1, J_2, J_3$ are Hermitian matrices of dimension (2j+1) where $j = 0, 1/2, 1, 3/2, 2,5/2,...$. Our task is to find, for each j, a matrix H that satisfies the requirements of (\ref{b6}).
 In the standard representation, $J_3$ is a diagonal matrix with the entries $m = j, j-1, j-2, ... -j$. The matrices $J_1, J_2$ have nonzero elements only one-off from the diagonal,
 \begin{equation}
J_{\pm} |j,m> = (J_1 \pm i J_2) |j,m> = \sqrt{(j\mp m) (j \pm m +1)} |j, m \pm 1>.\label{c1}
\end{equation}
Since H should commute with $J_3$, we take it as a diagonal matrix of dimension (2j+1).Combining these matrices we see,
\begin{eqnarray}
<j ,m\pm 1|J_{\pm} H |j, m> = <j, m\pm 1|J_{\pm} |j, m> H_{m,m}, \label{c2}\\
 <j, m\pm 1|HJ_{\pm}  |j, m> = H_{m\pm1 ,m\pm 1}<j ,m\pm 1|J_{\pm} |j, m>; \label{c3}
 \end{eqnarray}
And thus, requiring $H_{m,m} = -  H_{m\pm1, m\pm 1}$, one finds the simple solution that anti-commutes with $J_1, J_2$
\begin{equation}
H_{m,m'} = \delta_{m,m'}\;(-1)^{j-m}.\label{c4}
\end{equation}
In the case of (2j+1) even, this matrix H is just the repetition of the Pauli matrix $\sigma_3$, originally found.

\section{More Matrices like H}
For the general spin j, we have the matrices $J_1, J_2, J_3$ of dimension (2j+1) with indices $m = j, j-1, j-2, ... -j$. We found, above, a matrix H that obeyed, 
\begin{equation}
H J_3 = + J_3 H, \;\;\; HJ_1 = -J_1, \;\;\; H J_2 = - J_2 H; \;\;\;\;\; H_{m,m'} = \delta_{m,m'} (-1)^{j-m}.\label{d1}
\end{equation}

I will now rename this matrix $H$ as $H_3$ and ask if I can find two other matrices, $H_1$ and $H_2$, such that I have a system as follows.
\begin{equation}
H_i J_i = + J_i H_i \;\;no \; sum, \;\;\;\;\;\; H_i J_k = - J_k H_i, \;\; i \neq k = 1,2,3.\label{d2}
\end{equation}
For 2j odd, here is the solution, in terms of Hermitian matrices with square 1.
\begin{equation}
(H_1)_{m, m'} = \delta_{m',-m}, \;\;\;\;\; (H_2)_{m,m'} =-i \delta_{m', -m} (-1)^{j-m}.\label{d3}
\end{equation}
In the case of $j=1/2$ these three H matrices are just the Pauli spin matrices; and for any j the internal algebra of these H matrices looks just like the algebra of the Pauli matrices.
\begin{equation}
H_1 H_2  = -H_2 H_1 = iH_3, \;etc.\label{d4}
\end{equation}

The matrix $H_1$ is simply a string of ones along the "other diagonal" (upper right to lower left) of the matrix. (What we call the unit matrix $I$ is a string of ones along the "usual diagonal" (UL - LR) of the matrix. 

We can define the modified rotation operators, ${\cal{J}}_i = H_i J_i $ (no sum); and find the algebra,
\begin{equation}
{\cal{J}}_1\;{\cal{J}}_2 + {\cal{J}}_2\;{\cal{J}}_1 =  {\cal{J}}_3, \; etc.\;\;\;\;\; 
\end{equation}

Let us see some properties of $H_1$.

For an arbitrary square matrix A, indexed as above, we have the conventional transpose defined as $(A^T )_{m,m'}= A_{m', m}$. Visually, this operation is a reflection of the matrix about the usual diagonal. We can define a different operation, which is a reflection about the other diagonal,
\begin{equation}
A \rightarrow \;^TA, \;\;\;\;\; (\;^TA)_{m, m'} = A_{-m', -m}.\label{d5}
\end{equation}
We find that the matrix $H_1$ can produce this pair of operations,
\begin{equation}
H_1 A H_1 = \; ^T A ^T = \;^T(A^T) = (^TA)^T.\label{d6}
\end{equation}
We also find the identities
\begin{equation}
^TA = H_1 A^T H_1, \;\;\;\;\; ^T(AB) = (^TB)\;\;(^TA) .\label{d7}
\end{equation}
Furthermore we can have two other reflections on the matrix A: $H_1A$ is a reflection about the middle horizontal line; $AH_1$ is a reflection about the middle vertical line.

\section{Examining j=5/2}

We may first note that for j=3/2, we have 4x4 matrices for the $J_i$ and the larger algebra formed by making symmetrized products can be written in a block form of 2x2 matrices that can be expressed as the Pauli matrices $\sigma_1, \sigma_2, \sigma_3$. 

For the case of j=5/2, we have the three generating matrices $J_1, J_2, J_3$ as 6x6 Hermitian matrices. There is an additional symmetry in those matrices that can be seen by reflecting the matrix elements across the "other diagonal", which goes from upper-right to lower-left corners. This leads us to picture the 6x6 matrices in two-by-two block form, with each block being a 3x3 matrix. We focus first on the 3x3 matrix that sits in the upper-left quadrant.

For spin 1/2 we are familiar with the Pauli matrices and their very simple algebra: there are the three linear elements, $\sigma_1, \sigma_2, \sigma_3$, but any bi-linear forms are reducible to multiples of the unit matrix $I$ and those linear elements. For the generating matrices of any higher spin the algebra is more complex. 

For the matrices of j=5/2, we have the three linear elements, as noted, and then we have six bi-linear combinations: $J_i^2, i=1,2,3$ and $((J_i J_k)), i \neq k$ , where $((AB)) \equiv (AB + BA)/2$. That makes a total of 9 Hermitian matrices; and we know there is one combination that is a multiple of the unit matrix, $J_1^2 + J_2^2 + J_3^2 = j(j+1)I$. That leads us to consider a set of eight traceless Hermitian matrices of dimension 6x6. We wonder if the selected 3x3 matrices described two paragraphs above might have some relation to the matrices of the Lie algebra for SU(3), which is so famous in elementary particle physics.

Doing some tedious numerical calculations we find the answer to be, Yes. Here is the equivalence, relating these matrices for j=5/2 to the Gell-Mann matrices $\lambda_{1,2,...8}$ for the SU(3) Lie algebra.
\begin{eqnarray}
((J_1J_2)) \propto \lambda_5, \;\;\;\;((J_3J_1))-2J_1 \propto \lambda_6, \;\;\;\; ((J_3J_1))-J_1 \propto \lambda_1,\label{e1}\\
J_1^2 - J_2^2 \propto \lambda_4, \;\;\;\;((J_3J_2))-2J_2 \propto \lambda_7, \;\;\;\; ((J_3J_2))-J_2 \propto \lambda_2.\label{e2}
\end{eqnarray}
The symbols $\propto$ means there is a real(nonzero) multiplicative constant in this equality. The diagonal matrices $J_3,\; J_3^2$, adjusted to have zero trace, can be taken in two linear combinations to yield the two diagonal Gell-Mann matrices, $\lambda_3, \; \lambda_8$.

Are these results for spin $5/2 \rightarrow SU(3)$ new or have they been noted before? I do not know. I expect there may be a more sophisticated way to derive these mathematical results. Now I return to physics.

\section{Dirac-like Equations}

There is a long history of formulations of Dirac-like equations for particles of spin greater than 1/2. 
A paper by H. L. Baisya \cite{HLB} published in 1995 cites much of that literature and presents a detailed formulation for the case of $j=5/2$. That author follows a traditional approach for ordinary particles, using matrices of dimension 56x56 to achieve a Dirac-like equation that has three different masses. The related Klein-Gordon wave equation has up to six powers of the space-time derivatives. A more recent paper \cite{MMB} gives a similar close look at the case $j = 3/2$ and finds a fourth order K-G equation with two different masses.

That counting of masses and derivatives may be anticipated from the matrix identity, for any representation of the generators of O(3),
\begin{equation}
\prod_{m=-j}^{j} [\textbf{u} \cdot \textbf{J} -m] = 0,\label{f1}
\end{equation}
where $\textbf{u}$ is any unit vector in 3 dimensional Euclidian space. In the case of 2j odd, this equation lets one express the inverse of the matrix $U \equiv \textbf{u} \cdot \textbf{J}$. For example, for j=5/2, we have,
\begin{equation}
(U^2 - (1/2)^2) (U^2 - (3/2)^2)(U^2 - (5/2)^2)= 0, \label{f2}
\end{equation}
which we can rewrite as
\begin{equation}
U^2 [U^4 + a U^2 + b I] = cI,\label{f3}
\end{equation}
where $I$ is the (2j+1)-dimensional unit matrix. The expression in square brackets is thus proportional to the inverse $U^{-2}$. Note that this does not work for the case of 2j even, and that is because $J_3$ has an eigenvalue equal to zero.

It appears that Baisya's formulation could be used for tachyons, replacing his real mass parameter by an imaginary quantity. However, I wish to propose an alternative formulation that is much smaller.  I start by looking at the usual Dirac equation for spin 1/2, using matrices of dimension 4x4, written as 2x2 matrices of 2x2 matrices.
\begin{equation}
D \Psi = \left (\begin{array}{cc}
\omega &-\mathbf{k}\cdot \sigma \\
\mathbf{k} \cdot \sigma & - \omega
\end{array} \right) \Psi = \mu \Psi,\label{f4}
\end{equation}
where 2x2 or 4x4 unit matrices are implied and we see the familiar Pauli spin matrices $\sigma$.
We readily see, $D^2\Psi = (\omega^2 - k^2)\Psi = \mu^2\Psi,$ which is the Klein-Gordon equation.

I want to extend this for tachyons of any spin j = integer + 1/2. A particular choice of variables, that I have used in earlier studies of tachyon kinematics, involves separating the momentum vector into its magnitude and direction.
\begin{equation}
\textbf{k} = k\; \hat{k}.\label{f5}
\end{equation}
Next I introduce the (2j+1)-dimensional matrices for spin j
\begin{equation}
U = \hat{k} \cdot \textbf{J}.\label{f6}
\end{equation}
The extended Dirac equation (which I will "derive" a bit later) will work on a spinor $\Psi$ of dimension 2(2j+1) as follows.
\begin{equation}
D \Psi = \left (\begin{array}{cc}
\omega U&-k U \\
k U & - \omega U
\end{array} \right) \Psi = i\mu \Psi,\label{f7}
\end{equation}
where I have explicitly put the i in the mass term. Squaring this leads to the Klein-Gordon form,
\begin{equation}
D^2 \Psi = U^2 (\omega^2 - k^2)\Psi = -\mu^2 \Psi.\label{f8}
\end{equation}
Dividing by $U^2$ gives us the mass matrix,
\begin{equation}
M^2 =-\mu^2 U^{-2} \label{f9}
\end{equation}
and we know the (j+1/2) different eigenvalues of $U^{-2}$: $(j)^{-2}, (j-1)^{-2}, (j-2)^{-2}, ...(1/2)^{-2}$.

This looks almost too easy. In truth, I constructed the above form of $D$ in order to get this nice-looking result.\footnote{One may note that for j=1/2 Eq. (\ref{f7}) does not look like the standard Dirac equation Eq. (\ref{f4}); but it is equivalent to it.} Baisya gets the same result after a lot of hard work. In this spirit of easy model making one could add a factor U to the right hand side of Eq. (\ref{f7}) and thus eliminate the multiple masses; but that seems uninteresting. See Appendix A for some further discussion of this and alternative models.

One may complain that the matrix U is not a local operator in coordinate space since the usual quantum rules are for the full vector, $\hbar \textbf{k} = \textbf{p} \rightarrow -i \hbar \nabla$, and we take only the unit vector $\hat{k} = \textbf {k} / k$. I would respond to that criticism by saying we usually work in momentum space anyway; if we want space-time coordinates we just append $exp(-ik_\mu x^\mu)$ to the momentum space wave-function. Furthermore, for tachyons we know that the magnitude of the momentum $k = |\textbf{k}|$ is bounded away from zero; but for ordinary particles this could pose additional problems.

Here is a way to develop the Dirac-like equation given above. We start by writing down something that looks like a representation of the wavefunction for a tachyon of spin j:
\begin{equation}
\psi(x)=|\hat{k}, m> e^{i\textbf{k}\cdot \textbf{x} - i \omega t},\label{f10}
\end{equation}
where $|\hat{k}, m>$ is an eigenfunction of the (2j+1) dimensional matrix $U = \hat{k} \cdot \textbf{J} $ with eigenvalue m. The Dirac equation should be linear in space and time derivatives and also in some matrices that act algebraically in some finite vector space (representing the internal degrees of freedom).
Let's think at first of the spatial rotations, a sub-group of the Lorentz transformations. We know how to write a scalar product of vectors; so we start by writing,
\begin{equation}
-i \textbf{J} \cdot \nabla = -i(J_1 \partial_x + J_2 \partial_y + J_3 \partial_z).\label{f11}
\end{equation}
If this is one term in the new Dirac equation, we want to add another term, with $i\partial_t$ but it should multiply some (2j+1) dimensional matrix that is a scalar under 3-rotations. Here is our best guess,
\begin{equation}
i U \partial_t = i \hat{k} \cdot \textbf{J} \partial_t.\label{f12}
\end{equation}
For the familiar $j=1/2$ case, we know that we need to double the dimensions of the algebra. The reason for this is the fact that there are two representations of the full Lorentz group: one realized by the representations $K_i = +i J_i$ and the other by $K_i = -i J_i$. (I hope the reader can bear with the symbol "i" being used in two very different ways in the same equation.) So here is our guess at the full structure of the Dirac equation for a tachyon of spin j.
\begin{eqnarray}
\Psi = \left(\begin{array}{c}
\psi(x) \\
\phi(x) \end{array}\right), \;\;\;\;\; D \Psi =i\mu \Psi, \;\;\;\;\;D =   i U \partial_t\; \;\gamma_0 -i \textbf{J} \cdot \nabla \; \;\gamma_{50} , \label{f13} \\
\;\gamma_0= \left( \begin{array}{cc}
+I & 0 \\
0 & -I
\end{array}\right ) ,
\;\;\;\;\; 
\;\gamma_{5} = \left( \begin{array}{cc}
0 & +I\\
+I & 0
\end{array}\right) ,
\;\;\;\;\; 
\;\gamma_{50} = \gamma_5 \gamma_0 =\left( \begin{array}{cc}
0 & -I\\
+I & 0
\end{array}\right) ,\label{f14}
\end{eqnarray}
where $I$ is the $(2j+1)\times(2j+1)$ unit matrix. This is the formula given earlier (\ref{f7}).

There is one fine point in this that I should clarify: it concerns the quantity $\hat{k}$. The Dirac equation is something written in coordinate space. So I should define an operator in coordinate space $\hat{p}$, such that
\begin{equation}
\hat{p} \; e^{i\textbf{k} \cdot \textbf{x}} = \hat{k}\;e^{i\textbf{k} \cdot \textbf{x}}\;\;\;\;\forall \; |\textbf{k}| >0 ; \label{f15}
\end{equation}
and then I can write $U = \textbf{J} \cdot \hat{p}$, which becomes $\textbf{J} \cdot \hat{k}$ when acting on a plane wave. A question arises when I also have the familiar operator $\nabla$ acting in coordinate space: do $\nabla$ and $\hat{p}$ commute with each other? As long as they both operate on a plane wave, the answer is, Yes. This same discussion applies to the operator H.

We find that this theory can be put into a Lagrangian formalism with the term,
\begin{equation}
\bar{\Psi} (D - i \mu) \Psi, \;\;\;\;\; \bar{\Psi} = \Psi^\dagger \gamma_{50},\label{f16}
\end{equation}
which looks like what we had earlier for j=1/2 tachyons.

Finally, we face the essential task of showing the explicit Lorentz covariance of this Dirac-like equation for tachyons of any spin j = integer + 1/2. I have tried and failed to construct a representation of the Lorentz transformation on this generalized Dirac equation of dimension 2(2j+1). I do not say that it is impossible; but there appears to be an easier way to proceed - by appealing to the Poincar\'e Group - once we admit multiple masses that depend on $m$, the helicity. This implies that the magnitude of the momentum $k$ varies with the helicity due to the definition $k^2 = \omega^2 + |M^2|$. 

The Poincar\'e Group includes the Lorentz Group along  with the four operators $P^\mu$ that generate space-time translations. Those four operators commute with one another and they are represented by the plane wave $exp(i\textbf{k}\cdot \textbf{x} -i \omega t)$. The J and K operators mix up the components of this 4-momentum operator; but there is one invariant (a Casimir operator): $P^\mu P_\mu \rightarrow (\omega^2 - k^2)$. An irreducible representation of the Poincar\'e Group will be a vector space in which this invariant has a single real number as its value. That number may be positive (we call these ordinary particles, with $v<c$), or zero (light, with $v=c$), or negative (tachyons, with $v>c$). Among sophisticated physicists it is said that one way of rigorously defining a subatomic particle is this: "Particles are at a very minimum described by irreducible representations of the Poincar\'e group."  (This quote found in the Wikipedia article on the Poincar\'e group.) 

Therefore, seeking \emph{irreducible} representations of the Poincar\'{e} group for tachyons obeying our modified Dirac equation, we would say that we have only to study \emph{distinct 2-dimensional vector subspaces}:  the pair of states in each having the same magnitude of helicity but distinguished by the sign of the helicity. Each such subspace now looks like the ordinary starting point for Dirac's original wave equation. We can define the Pauli matrices that work within each subspace; and then we can double that subspace to make the traditional Dirac spinors (4-components). The Lorentz transformations are built exactly as for the original Dirac equation. We can then (as done in my earlier papers) replace the real mass by an imaginary and insert a factor $\gamma_5$, next to the usual $\gamma_0$, in the definition of the adjoint Dirac spinor. (This leaves open the question of whether to insert a further $\pm$ sign in the Lagrangian for each subspace, which could be interpreted as a relic of the big matrix H introduced in the early sections of this paper.) 

In the case of j=1/2, we identified the two helicity states ($m= + 1/2$ and $m=-1/2$) with the names "particle" and "antiparticle"; that same identification now appears proper for higher values of j. There is more one may do with this new construction but I shall stop for now. 

\section{Discussion}
In the earlier studies of j=1/2 tachyons, this metric operator H led to the most interesting conclusion that the states of opposite helicity (also identified as the particle and anti-particle) contributed to the Energy-Momentum tensor with opposite signs.\cite{CS2} It was necessary to use the Dirac equation (with $\mu \rightarrow i\mu$) to get most of the interesting results for the quantized tachyon spinor. 

The present work starts by finding the generalization of that metric H for all spins; and then we also find some provocative results for spin 5/2: an algebraic connection with SU(3).

The attempts to construct a Dirac-like wave equation for tachyons of any spin led to interesting challenges. Former work on such equations for ordinary ($v<c$) particles led to enormous dimensions of the representation; but for tachyons we found simplifications that led to much simpler models. With the acknowledgment of multiple mass states (for 2j odd) we finally found the irreducible subspaces of much smaller dimension and a complete resolution that seems largely a copy of the earlier results for j=1/2.

It may seem strange (unfamiliar) that we started with (2j+1)-dimensional matrices for the Lorentz Little Group but we ended up using just 2-dimensional subspaces for the final irreducible representations of the Poincar\'e Group. The original operators $J_i, K_i$ were viewed as dynamical variables but the 2-dimensional Pauli matrices $\sigma_i$ used later are merely convenient algebraic tools. One may ask, What is the role of the larger matrices? The operator $J_3$, or what we called U in the text above, is diagonal and its eigenvalues tell us the mass of each particle and also distinguishes between "particle" and "anti-particle". What about $J_1, J_2$? We did not use them to construct the Lorentz transformations of the new Dirac-like equation; they would appear to mix states of different magnitudes of the helicity. Thus they would produce "mass-mixing" of the particle states if those larger operators were used in constructing interactions between the free-tachyon fields (which are the focus of the present paper) and other fields. That is a ripe topic for future investigation. Again, the known physics of neutrinos seems resonant among these mathematical studies.

\vskip 1cm

{\bf Appendix A: Alternative Models and Data Fitting for Neutrino Masses}
The model given in Eq.(\ref{f7}) for the modified Dirac equation for tachyons gives a simple formula for the multiple masses, $M^2 = -\mu^2 U^{-2}$. How does this formula compare with the experimental data on the masses of the three types of neutrinos?  We take the case j = 5/2, which gives us three masses:
\begin{equation}
m_1^2 = -\mu^2 (1/2)^{-2}, \;\;\;\;\; m_2^2 = -\mu^2 (3/2)^{-2},, \;\;\;\;\; m_3^2 = -\mu^2 (5/2)^{-2}.
\end{equation}
The constant $\mu$ is just an arbitrary scale factor so we can look at the ratio,
\begin{equation}
R = (m_1^2 -m_2^2)/(m_2^2 - m_3^2).
\end{equation}
The best experimental data I could find yields $R = 32.4 \pm 1.1$; and the above formula gives $R=12.5$.
That is not a very good fit but it seems perhaps "in the right ballpark".

One can, of course, make alternative models for the tachyon Dirac equation with arbitrary functions of the matrix U. For example, if we get a mass formula $M^2 = -\mu^2 |U|^{-3}$, then this gives a theoretical value for $R =33.16$, which is within the error limits of the experimental data for neutrinos. An alternative model, with $M^2 = - \mu^2 (sinh \;\alpha U)^{-2}$, also fits that experimental data precisely with $\alpha = 3/2$.

\end{document}